\documentclass{aastex}
\usepackage{spr-astr-addons}
\usepackage{url}

\RequirePackage{color}

\begin{document}

\title{The Wide Field Spectrograph (WiFeS):\\ Performance and Data Reduction}
\shortauthors{Autors et al.}

\author{Michael Dopita, Jonghwan Rhee, Catherine Farage,  Peter McGregor, Gabe Bloxham, Anthony Green, Bill Roberts, Jon Nielson, Greg Wilson, Peter Young} 
\affil{Research School of Astronomy \& Astrophysics, \\ The Australian National University, Cotter Rd., Weston Creek, ACT 2611, Austalia}
\author{Peter Firth}
\affil{Department of Physics, The University of Queensland, \\ St. Lucia, QLD 4072, Australia}
\author{Gianni Busarello, Paola Merluzzi}
\affil{INAF-Osservatorio Astronomico di Capodimonte, \\ Salita Moiariello 16, Napoli 80131, Italy}
\email{Michael.Dopita@anu.edu.au}


\begin{abstract}
This paper describes the on-telescope performance of the Wide Field Spectrograph (WiFeS). The design characteristics of this instrument,  at the Research School of Astronomy and Astrophysics (RSAA) of the Australian National University (ANU) and mounted on the ANU 2.3m telescope at the Siding Spring Observatory has been already described in an earlier paper (Dopita et al. 2007). Here we describe the throughput, resolution and stability of the instrument, and describe some minor issues which have been encountered. We also give a description of the data reduction pipeline, and show some preliminary results.

\end{abstract}



\section{Introduction}

The recently constructed Wide Field Spectrograph (WiFeS) has been designed to operate on the ANU 2.3m telescope at Siding Spring Observatory and to deliver excellent throughput, spectral resolution, spectrophotometric stability and spatial and spectral stability over its full field of view. WiFeS  is an integral field spectrograph that draws on the heritage of the concentric image-slicing design concepts of the NIFS spectrograph built for for the Gemini North \citep{mcg99,mcg03}, and upon the experience of the Double Beam Spectrograph (DBS) \citep{Rodgers88} which WiFeS replaced.

WiFeS provides a $25\times38$~arc sec. field with 0.5~arc sec. sampling along each of twenty five $38\times1$~arc sec slitlets.  The output format is optimized to match the $4096\times4096$~pixel CCD detectors in each of two cameras individually optimized for the blue and the red ends of the spectrum, respectively. A process of ``interleaved nod-and-shuffle" is applied to permit quantum noise-limited sky subtraction. Using Volume Phased Holographic (VPH) gratings, spectral resolutions  of 3000 and 7000 are provided.  The full spectral range is covered in a single exposure in the $R=3000$ mode, and in two exposures in the $R=7000$ mode. The complete instrument description and the design considerations for WiFeS have been fully described in an  earlier paper \citep{dop07}. 
The different standard modes of operation of the WiFeS spectrograph are summarized in Table \ref{Table1}.
\begin{table*}[t]\label{Table1}
\centering
\caption{The standard WiFeS grating and dichroic sets and their wavelength coverage}
\begin{tabular}{@{}lcccccc@{}}
\tableline \tableline 
& & {\bf Blue} & & & {\bf Red} & \\
& \multicolumn{2}{c}{$R = 7000$} & {$R = 3000$} & \multicolumn{2}{c}{$R = 7000$} & {$R = 3000$} \\
\tableline
Grating  & U7000 & B7000 & B3000 & R7000 & I7000 & R3000  \\
Dichroic & RT480 & RT615 & RT560 & RT480 & RT615 & RT560  \\
$\lambda_{min}$ (\AA) & 3290 & 4180 & 3200 & 5290 & 6830 & 5300  \\
$\lambda_{0}$ (\AA) & 3850 & 4900 & 4680 & 6200 & 8000 & 7420  \\
$\lambda_{max}$ (\AA) & 4380 & 5580 & 5900 & 7060 & 9120 & 9800  \\
\tableline

\end{tabular}
\end{table*}

The WifeS instrument is designed as a facility instrument. Consequently, the science mission of the WiFeS spectrograph is very broad and encompasses observations of single stars,  clusters of stars, extended nebulosities, galaxies in the nearby and distant universe and gamma-ray bursts. The following list of titles drawn from the WiFeS Science Verification Mission gives an idea of the scope of the science which is enabled by the WiFeS instrument:
\begin{itemize}
\item{Solving the Chemical Abundance Controversy for HII Regions}
\item{Dynamics \& Abundances in Young SNRs}
\item{Physical Conditions in Externally Irradiated Protostellar Outflows}
\item{The Ejecta and the Structure of Eta Carinae}
\item{A complete spatial and dynamical study of the micro-quasar SS 433}
\item{Physical Conditions in Propylids of the Orion Nebula}
\item{Physical Conditions in Old Nova Shells}
\item{The Age-Velocity-Metallicity Relation near the Sun}
\item{Do globular cluster  RGB and AGB stars have dramatically different CN abundances?}
\item{Red-Giant Binaries in the LMC	}
\item{The Ages \& Metallicities of  LMC Globular Clusters}
\item{Supernova Remnants, HII Regions and the Abundance Gradient in M83}
\item{The nature of the Filaments of Cen A}
\item{The Structural \& Dynamical Parameters of Disk Galaxies	}
\item{Gas-Rich Dwarf Galaxies: Dark Matter \& the Metallicity Floor of the Local Universe}
\item{Galactic Winds: Just How Powerful are they?}
\item{The Circumnuclear Environment of a Sample of Nearby Seyfert Galaxies}
\item{Galaxy Mergers \& Metallicity}
\item{Luminous IR Galaxies: The WiFeS GOALS Survey}
\item{NGC 4696: Prototype of the interaction between a Radio jet and a Galaxian Medium.}
\item{Feedback mechanisms on galaxy formation: the constancy of jets in radio galaxies.}
\item{AGN Black Hole Masses \& the Cause of Accretion}
\item{ The transformation of galaxies in the Shapley supercluster}
\item{Assembly, Accretion and Outflows in High Redshift Radio Galaxies.}
\item{Prompt Spectroscopy of Gamma-Ray Burst Sources}
 \end{itemize}

 The WiFeS instrument was commissioned on the 2.3m telescope in March-April 2009, and functional verification occurred in April 2009. The purpose of this paper is to describe the performance characteristics as achieved on the telescope, draw attention to  the instrumental issues which have been identified, describe the characteristics of the data reduction pipeline, and to present some preliminary results obtained with the instrument. Further information on  observing with WiFeS, on the WiFeS data reduction procedure and on the telescope and instrument control software can be found at: \textsf {http://www.mso.anu.edu.au/observing \newline /ssowiki/index.php/WiFeS\_Main\_Page}.

\section{The on-telescope operation of WiFeS}
\subsection{Data Accumulation Modes}
The WiFeS instrument offers a number of different configuration and data accumulation modes of observation, designed to optimize the science data obtained on different classes of astronomical object, be they extended nebulosity and galaxies or stars. The fixed instrument configurations have already been listed in Table \ref{Table1}. 

In addition, the CCD can be set up for either single pixel binning or double binning in the spatial direction. The single binning mode gives a $4096\times4096$ data format, with 0.5 arc sec pixels in the spatial direction, and is used in conditions of excellent seeing where the full spatial resolution is required, and where the read-out noise is not a key issue. Alternatively, in the double binning mode we have a data format of $4096\times2048$ data format, with 1.0 arc sec pixels in the spatial direction. This has the advantage of doubling the signal count per pixel for the same read-out noise. It also reduces the read-out time by a factor two. In double binning, the instrument becomes sky-limited in the B3000 and R7000 modes, so this mode particularly well-suited for the observation of faint galaxies.

There are two principal data accumulation modes:
\begin{enumerate}
\item{{\bf ``Classical Mode''}, in which the data is simply accumulated in the red and/or blue cameras for a given exposure time \emph{and }}
\item{{\bf ``Nod-and-Shuffle Mode''}, in which both object and nearby sky-background data is accumulated on the CCD chips in the red and/or blue cameras for a given (on-source) exposure time. In this case the actual observation time is the sum of the on-source and off-source exposure time plus the overheads associated with manipulating the shutter, nodding the telescope, re-acquiring guide stars and shuffling the charge on the chip. However, the overhead can be largely eliminated when the object covers less than half of the observing aperture (such as when observing individual stars or small galaxies), and the observer can perform the variant of nod-and shuffle called  \emph{``Sub-Aperture Nod-and-Shuffle''}, described below. In nod-and-shuffle, the sky patch can in principle be chosen several degrees away. The offset between the object and its sky patch is limited only by the overhead induced by the slew time between the patches, and the need to look at a region of the sky which has a very similar OH night-sky spectrum to the region being observed in the primary beam. In practice the separation is usual chosen to be a few arc min. on the sky.}
\end{enumerate}

\subsection{Classical Mode with Equal Exposures}\label{equal}
This is the simplest mode of operation in which shutter is opened for a given exposure time in both cameras, and then the accumulated charge on the CCD is read out in the normal manner. The total observation time is then the sum of the expose time plus the readout time -- which is approximately 45 sec. In this mode only half the chip contains science data.

\subsection{Classical Mode with Unequal Exposures}\label{unequal}
This mode is used when an object is much brighter in either the blue or red (usually the red), or where there is a very bright emission line in either the blue or the red (usually red for arcs, but sometimes blue for objects), and one also wishes to achieve high dynamic range. In this case, the observer splits up the exposure with the bright line or continuum between a number of shorter (unsaturated) sub-exposures, while continuing with a long single exposure with the other camera. In this mode the observational sequence is as follows:
\begin{enumerate}
\item{Open shutter, expose for $x$ seconds.}
\item{Close shutter, read out CCD which requires the short sub-exposure and save file.}
\item{Repeat steps (1) and (2) up to N times, then read out both CCDs and save the files.}
\end{enumerate}

We now have one exposure (of $Nx$ seconds on-target exposure time) recorded in one arm, and $N$ sub-exposures, each of $x$ seconds on-target exposure time recorded for the other arm. The total observational time will be $(Nx + Nr)$ sec, where r is the read-out time (about 90 sec.). This total observational time will be limited to about 60 min by dark current and cosmic ray considerations. Therefore, for 10 sub-exposures, each would have to be no longer than about 240 sec, giving an maximum observational efficiency of 66\% in this mode. 

 \subsection{Nod-and-Shuffle Mode with Equal Exposures}\label{NS_equal}
In this mode the telescope control system (TCS) and instrument control system operate together, as follows:
 \begin{enumerate}
\item{Acquire science target, open the shutter and expose for $x$ seconds.}
\item{Close the shutter, shuffle the charge by 80 pixels to place the charge in the un-illuminated space between the images of the slitlets, suspend the guiding on the offset guide star, and nod the telescope to point at a region of sky which can be used for sky reference. One can also acquire a separate guide star for the sky position if desired.}
\item{Open the shutter and expose for $x $ seconds.}
\item{Close the shutter, shuffle the charge back by the same number of pixels, in order to return the signal accumulated on the object to its original position on the CCD, nod the telescope back to the nominal coordinates of the object, and re-activate the guiding on the offset guide star.}
\item{Repeat steps (1) to (4) for a further $N-1$ cycles. }
\item{Read out both CCDs and store the images.}
\end{enumerate}

Not counting the acquisition time, the total observation time will be $(2Nx + 2Nt + Na + r)$ sec. and the effective on-object exposure time will be $Nx$ sec, where  $x$ is the exposure time in the individual sub-exposures, $t$ is the telescope nod time, $a$ is the guide star re-acquisition time and $r$ is the CCD readout time. Typical values for an observation could be $N = 10, x = 150, t = 3, a = 6,$ and  $r = 90$, giving a total (on object) observation time of 1500 sec, a total observation time of 3210 sec. and an observational efficiency of  47\%. This might seem low, but this is compensated by the very high quality of the sky subtraction. This enables many such observations to be co-added with shot-noise image statistics. 

In some cases, observers will use blank-field searches for very faint objects such as Lyman Break objects at redshifts greater than $z = 2$. In this case, both fields contain sky as well as the searched-for objects. Objects appearing in the ``object'' aperture will appear as positive signals in sky-subtracted images, and objects appearing in the ``sky'' aperture will give negative signals, thus allowing one to be distinguished from the other.

 \subsection{Nod-and-Shuffle Mode with Unequal Exposures}\label{NS_unequal}
 This is a more efficient mode of providing data in which the sky and the object counts are accumulated together. It is essentially identical in operation to the previous mode, except that the integration timescale in the sky is shorter, typically 50\% of the on-object exposure time. Not counting the acquisition time, the total observation time is $(Nx +Ny + 2Nt + Na + r)$ sec, where the symbols have the same meaning as in the previous section, and the sky exposure time is now $x$ sec. For $N = 10, x = 200, y=100, t = 3, a = 6,$ and  $r = 90$ the total observation time is still 3210 sec, but the on-target observational efficiency is now increased to 62.3\%.
 
 In practice, the shorter sky observation can be compensated for by summing the sky data in the sky direction over two pixels. The signal to noise  in strong sky lines is then increased by $\sqrt 2$, so entirely compensating for the shorter exposure time in the sky reference.

 \subsection{Sub-Aperture Nod-and-Shuffle Mode }\label{SA_NS_equal}
This mode is the recommended one for single star observing,  or in the case where we are dealing with objects such as distant galaxies in which the object does not fill more than half of the WiFeS aperture ($25\times19$ arc sec.). It is functionally indistinguishable from the previous nod-and-shuffle mode with equal exposures (section \ref{NS_equal}), but, since the science object is being observed at all times, the sub-aperture nod-and-shuffle mode offers much higher observational efficiency. 
 
In this mode, we organize that the star or galaxy is centered in the middle of one half of the WiFeS aperture (at an offset of $x=12.5; y= 9.5$ arc sec. from a chosen corner of the WiFeS field. We then arrange to nod the telescope along the long axis of the WiFeS field so as to place the star at field coordinates $x=12.5; y=28.5$ arc sec. following nod. This has the advantage that the object is always in the aperture while the CCD is being exposed, giving an effective on-target efficiency of $2Nx/(2Nx + 2Nt + Na + r)$. This doubles the efficiency, (to 93.4\% using the figures given in the example given in the earlier section (\ref{NS_equal}).
 
 In the extracted (sky-subtracted) image, we extract a (positive) signal at $x=12.5; y= 9.5$ arc sec. from a corner of the WiFeS field, and a (negative) signal from coordinates  $x=12.5; y=28.5$ arc sec., and then add the moduli of these signals to obtain the final sky-subtracted spectrum.

\begin{figure*}[t]
\centering
\includegraphics[width=\textwidth]{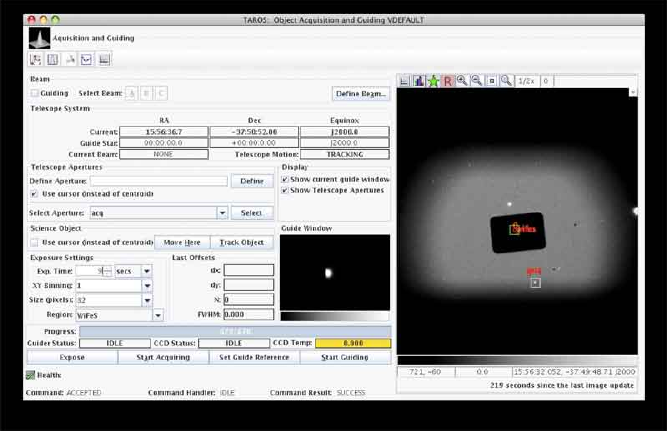}
\caption{An example of the TAROS acquisition and guide GUI. In this example, a science object has been placed in the WiFeS science aperture, an offset guide star has been selected (marked as a box below the aperture), and the system is actively guiding and updating the guide star image in the selected guide window.}\label{Guide-GUI}
\end{figure*}

\section{Telescope \& Instrument Control System}
All of the telescope control and observing functions described above are controlled through the software developed by the Research School of Astronomy and Astrophysics for this purpose. The software is called the Telescope Automation \& Remote Observing System (TAROS). Currently, TAROS is used to remotely observe or observe on-site with the 2.3m telescope and to provide automated, queue-scheduled operation of the SkyMapper
telescope, a wide-field broad-band fast imaging telescope currently being commissioned at Siding Spring \citep{keller07}. The instruments available via TAROS at the 2.3m are WiFeS, the Imager and the Echelle spectrograph.

TAROS is comprised of several sub-systems which handle such things as detector control, acquisition and guiding, instrument component control, telescope control, data archiving and communications. All data taken with TAROS is archived and a database is used to store information about each observation.

The TAROS software operates on a master / slave processing basis. The master program controls all the slave processes to achieve time synchronization of the observing operation. The master program monitors the health and status of all sub-systems, and reports these to the observer as status flags or as fault reporting. For both remote and on-site observing, a Java user interface has been developed. For automated observing, an observation scheduler has been developed in a way that allows specialised schedulers to be developed to suit individual observing programs.

The TAROS Graphical User Interface (GUI) is a Java application which runs under a Java Runtime Environment (JRE) of at least version 1.5. It will run on any modern operating system. It has been developed and tested on Mac OSX (10.4 and 10.5), Linux (Fedora Core 5, Ubuntu 8.04) and Solaris 10 (x86 and SPARC). It requires at least 300MB RAM and the recommended minimum connection speed is 10~Mbps. 

The TAROS GUI provides real-time meteorological data from the local weather station, a real-time all-sky image provided by the University of New South Wales and weather maps from the Australian Bureau of Meteorology. It also provides complete control of the telescope dome, telescope functions, offset guiding functions, and instrument control at both an instrument and instrument component level. Finally, it also provides telescope and instrument health and status reporting. An example of one of the TAROS GUIs, the acquisiton and guide window,  is shown in figure \ref{Guide-GUI}.

With the TAROS software safe remote observing is possible from any site which has a sufficiently fast network connection. To facilitate this mode of operation, TAROS offers a full control of the level of image compression used by the image displays in the WiFeS and acquisition and guide windows.

\begin{figure}[t]
\centering
\includegraphics[width=0.45\textwidth]{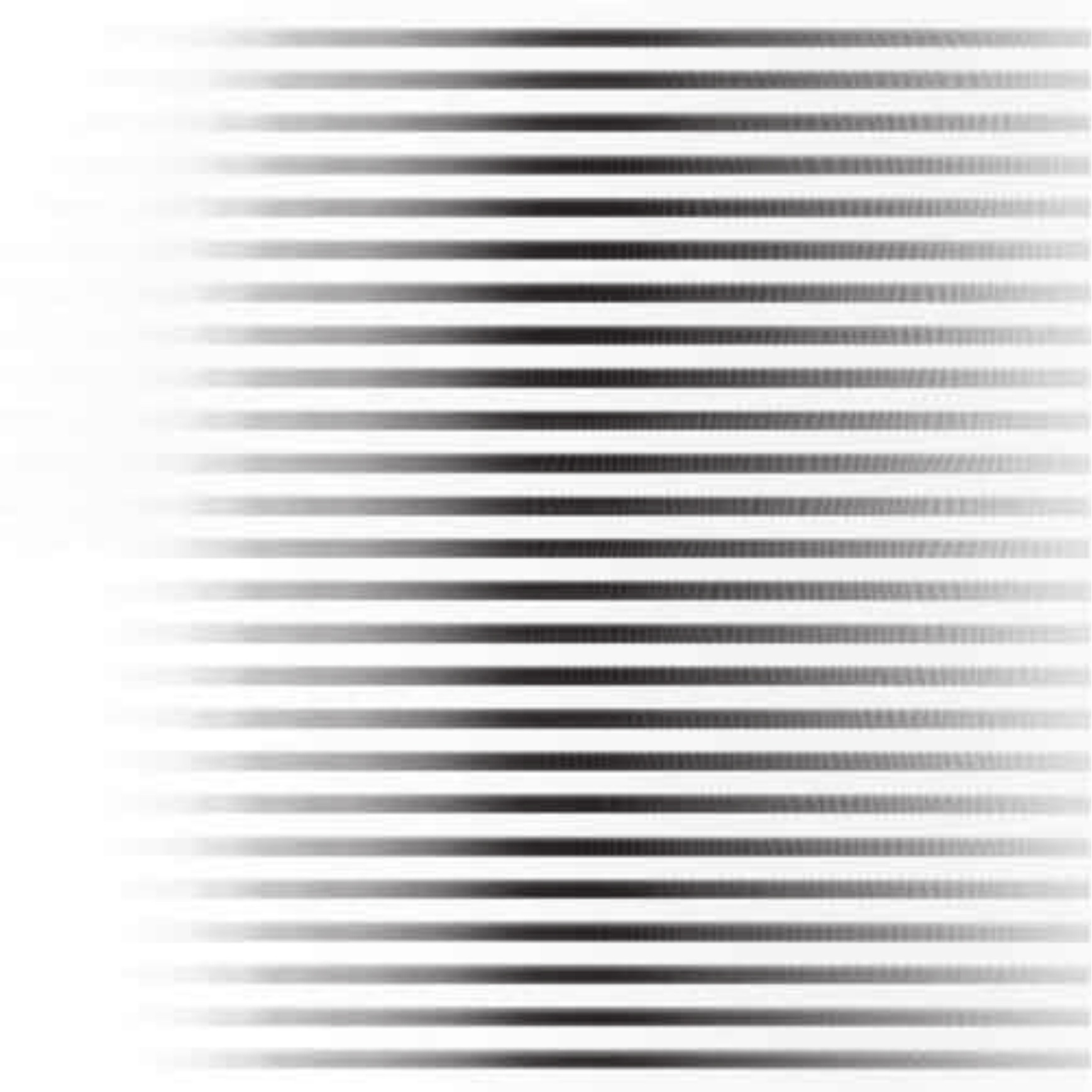}
\caption{A flat field image obtained with a quartz-iodine (QI) lamp in the R3000 mode showing the format of the raw WiFeS data. The stripes are the primary spectra of the 25 slices, and the empty area is the slightly wider interleaved area provided for the nod-and-shuffle mode. Note the CCD fringing visible at long wavelengths (right-hand side), which peaks at about 20\%. The dark region on the left-hand side is the consequence of the cut-off in the transmission of the dichroic below $\sim 580$~nm.}\label{format}
\end{figure}

\section{WiFeS data reduction pipeline}

The raw data format of the WiFeS instrument is shown in figure \ref{format}. This consists of 25 long-slit spectra corresponding to each 1.0 arc sec. wide slice produced by the image slicer. Each spectrum covers 76 pixels on the detector, and they are separated by 160 pixels to allow a sufficient space between them for nod-and-shuffle operation. There are thus a total of 1900 spaxels (\emph{i.e.} the spatial pixels of a spectral datacube). Each spectrum is 4096 pixels long. We have two detectors operating together, so, for each exposure we obtain two independent spectra. The objective of the reduction package is to convert this data into a cosmic-ray cleaned, bias-subtracted, flat-fielded, wavelength and flux calibrated three dimensional data cube. From such data cubes, the spectra of individual objects can be extracted, monochromatic images, line-ratio or line index images can be produced, and radial velocity maps or velocity dispersion maps of science objects can be derived.

The WiFeS data reduction program is based on the NOAO IRAF software. The WiFeS data reduction package package has been developed from the Gemini IRAF package used for reduction of the Near-IR Integral Field Spectrograph (NIFS) data \citep{mcg03}, since the data format and the steps needed to transform raw data to a calibrated data cube are quite similar in both instruments. Thus several tasks from the Gemini IRAF package are invoked during the WiFeS data reduction.  The WiFeS data reduction pipeline consists of four primary tasks - {\bf wifes, wftable, wfcal, wfreduce}. 

\begin{description}

\item[\bf wifes] executes the WiFeS package and sets up  the WiFeS package enviroment parameters related to the locations and names of raw, reduced, and calibration data files. 

\item[\bf wftable] converts raw data in a single-extension FITS file format to  the Multi-Extension FITS (MEF) file format. Another function of this task is  to create text files including list of files based on data type such as bias, flat, arc, wire, sky, and object. These lists of files are used during subsequent procedures, for instance, {\bf wfcal} and {\bf wfreduce}.

\item[\bf wfcal] identifies and processes the set of standard calibration frames. The required calibration frames are the bias, the flat-field obtained by diffuse illumination of the science aperture with a quartz-iodine (QI) lamp, ``wire'' frames obtained in the same way, but using the coronagraphic aperture, and finally, arc lamp data frames. All these frames  are required to reduce the science data. These calibration frames may be generated either using the QI lamp internal to the spectrograph, or dome illumination via a set of external QI lamps on the top-end ring of the telescope. The second method is preferred because the internal lamps are found to produce artifacts caused by uneven illumination of the science aperture.

\item[\bf wfreduce] applies the basic calibration to science object images. It also does sky subtraction, telluric feature correction and flux calibration,using the calibration solutions obtained from {\bf wfcal}.

\end{description}
\begin{figure}[t]
\centering
\includegraphics[width=0.45\textwidth]{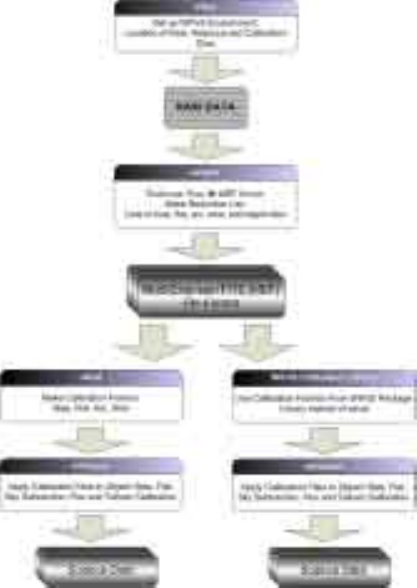}
\caption{The WiFeS data reduction procedure.}\label{Flow_Chart}
\end{figure}

The WiFeS data reduction can be performed in two ways, depending on which specific procedure is required. The logical flow chart is shown in figure \ref{Flow_Chart} The normal method is to carry out all procedures---
{\bf wftable}, {\bf wfcal}, and {\bf wfreduce}. The other method is to skip the {\bf wfcal} procedure that makes the calibration frames and use the stored calibration library files instead. This takes advantage of the fixed-format modes of the instrument.

\begin{figure*}[t]
\centering
\includegraphics[width=1.0\textwidth]{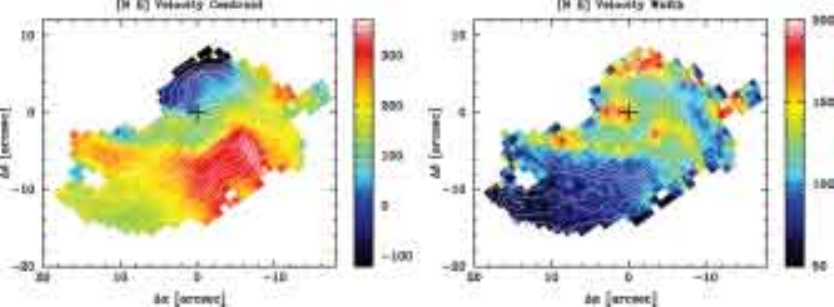}
\caption{Gas velocity and velocity dispersion maps extracted from the WiFeS data. These images show the [N~II] $\lambda 6484$\AA\ data for the first-ranked cluster Elliptical galaxy NGC~4696 (from Farage et al. (2010), in preparation. The left-hand panel is the radial velocity of the gas measured with respect to the systemic velocity (km s$^{-1}$), and the right-hand panel shows the local velocity dispersion of this gas (also in km s$^{-1}$). On each of these images are superimposed contours of the [N~II] line intensity.}\label{NGC4696}
\end{figure*}

To demonstrate the kinds of data products which can be extracted from the WiFeS reduced datacubes,  we provide two examples. In figure \ref{NGC4696} we show an example of both the radial velocity and the velocity dispersion for [N~II] emitting gas in this elliptical galaxy. In these images, the intensity of the emission line is indicated as contours. For those interested in making the comparision of these images with other data, this galaxy has been studied in depth by \citet{crawford05}, who present a set of images (in unsharp-masked H$\alpha$ + [N~II], radio brightness, X-ray emission and a reddening map) all at a similar scale to figure \ref{NGC4696}.

\begin{figure}[t]
\centering
\includegraphics[width=0.4\textwidth]{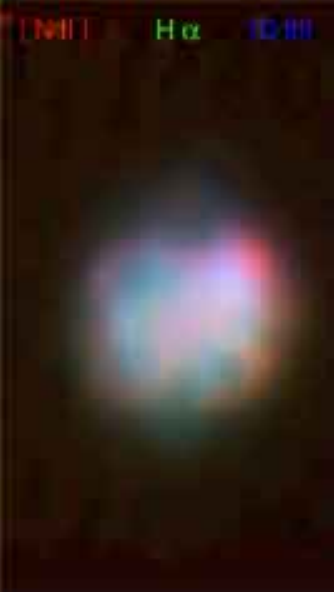}
\caption{WiFeS used as a multi-band imager. This is a reconstructed three-color [NII] 6484\AA\, H$\alpha$ and [O III] 5007\AA\ image of the famous planetary nebula NGC~3918 showing its ansae, as well as the complex point-to-point behavior of the excitation conditions within it. This image has been boxcar smoothed by the pixel resolution in both directions to remove the steps in intensity which would otherwise occur at pixel boundaries.}\label{NGC3918}
\end{figure}

 Spectral slices taken from WiFeS data cubes can also be used to construct multi-colour images. As an example, in figure \ref{NGC3918}, we show a three-color image of the Planetary Nebula NGC~3918 extracted from the datacube. This has been magnified and then box-car smoothed by the pixel resolution in both directions so as to remove the discontinuities introduced by the image pixelation. Figure \ref{NGC3918} demonstrates how the instrument can be used as a direct imager. Images such as this could not only be generated in emission lines, but also in absorption lines, line ratios or metallicity indices. An image such as figure \ref{NGC3918} would normally be obtained by first observing with three different emission line interference filters, then in the adjacent continuua, and finally processing the resulting images to generate a continuum subtracted emission line color map. A single WiFeS data cube provides the material for many such images which can be used for many astrophysical purposes.
\begin{table*}
\centering
\caption{Overview of WiFeS instrument performance characteristics.}\label{Table2}
\begin{tabular}{@{}ll}
\tableline
{\bf Spectral Resolution} (R=$\lambda/\Delta\lambda$):&  \\
\hspace{1em}High Resolution & Achieved: R=6800 (Velocity Resolution 50 km/s) \\
\hspace{1em}Low Resolution & Achieved: R=2900 (Velocity Resolution 105 km/s)\\
{\bf Spectral Coverage:} & \\
\hspace{1em}High Resolution & 329--558 nm (Blue)\hspace{1.5em}529--912~nm (Red)
\\
\hspace{1em}Low Resolution & 329--590 nm (Blue)\hspace{1.5em}530--980~nm (Red)
\\
{\bf Field of View:} & 25$\times$38 arcsec \\
{\bf Detector:} & Fairchild Coated CCDs optimized for both Blue and Red channels.\\
\hspace{1em} Format & 4096$\times$4096 pixels CCD \\
\hspace{1em}Gain & 0.9 e$^{-}$/ADU \\
\hspace{1em}Readout Noise & 5.0 e$^{-}$ \\
\hspace{1em}Dark Current & 3.8 e$^{-}$/hr \\
\hspace{1em}Pixel Size & 15$\mu$m square \\
\hspace{1em}Pixel Scale & 0$\farcs$5/pixel \\
\hspace{1em}Slice Spatial Width & 1$\farcs$0 wide (2 pixels on the CCD) \\
{\bf Limiting Magnitude:} & \\
\hspace{1em}Stellar Source & $\sim$21.5 mag \\
\hspace{1em}Extended Source & $\sim$10$^{-17}$ erg/cm$^{2}$/arcsec$^{2}$/\AA/s
(Surface Brightness) \\
\tableline
\end{tabular}
\end{table*}
\newpage

\section{Measured performance  of WiFeS}

\subsection{General}
The overall performance characteristics of the WiFeS instrument are laid out in Table \ref{Table2}. Note in particular that the achieved spectral resolution is very close to the theoretical limits imposed by the width of the entrance slices; 1.0 arc sec. on the sky, or 30$ \mu$m (two pixels) on the detector. 

\subsection{Spectral Resolution}
The measured width of arc spectrum lines is 2.2 pixels on the detector with the instrument in correct focus, with a variation of less than 5\% across the face of the detector. This is consistent with the expected optical performance and error budget. This showed that the contributions to the width of spectral lines should be, in order of size, the entrance slit, the spectrograph aberrations (which are primarily high order astigmatism from the collimator), and the finally, the camera optics (\emph{see} figure 13 of \citet{dop07}). The camera aberrations were measured in isolation on an optical bench subsequent to the assembly of the cameras, and these were found to be typically less than 8$\mu$m at the detector, again as designed.

It should be noted that the spectrograph was designed in such a way that the field distortion produced by the spectrograph is balanced against the (opposite) field distortion produced by the cameras. Thus, with a suitable roll adjustment of the cameras, we can contrive to have spectra which disperse along a single row of the detector. Tests performed during the commissioning confirm this to be the case. This much simplifies the subsequent extraction of the spectra. However, a seventh order polynomial is required in order to produce a sufficiently accurate ($\leq 0.05$ pixel) wavelength solution along the dispersion direction.

\subsection{Spectrograph Stability}
Since the spectrograph cameras do not have built-in thermal regulation, we might expect some drift in the spectra during the night. The cameras are designed to be fully compensated for both focal plane scale and focus as a function of temperature (focal plane shift less than 0.2 pixels for a temperature change of 5C) so this effect is too small to be measurable. However, it is impossible to compensate for the change of grating pitch as a function of temperature, which effectively changes the dispersion of the spectra. 

\begin{figure}[t]
\centering
\includegraphics[width=0.45\textwidth]{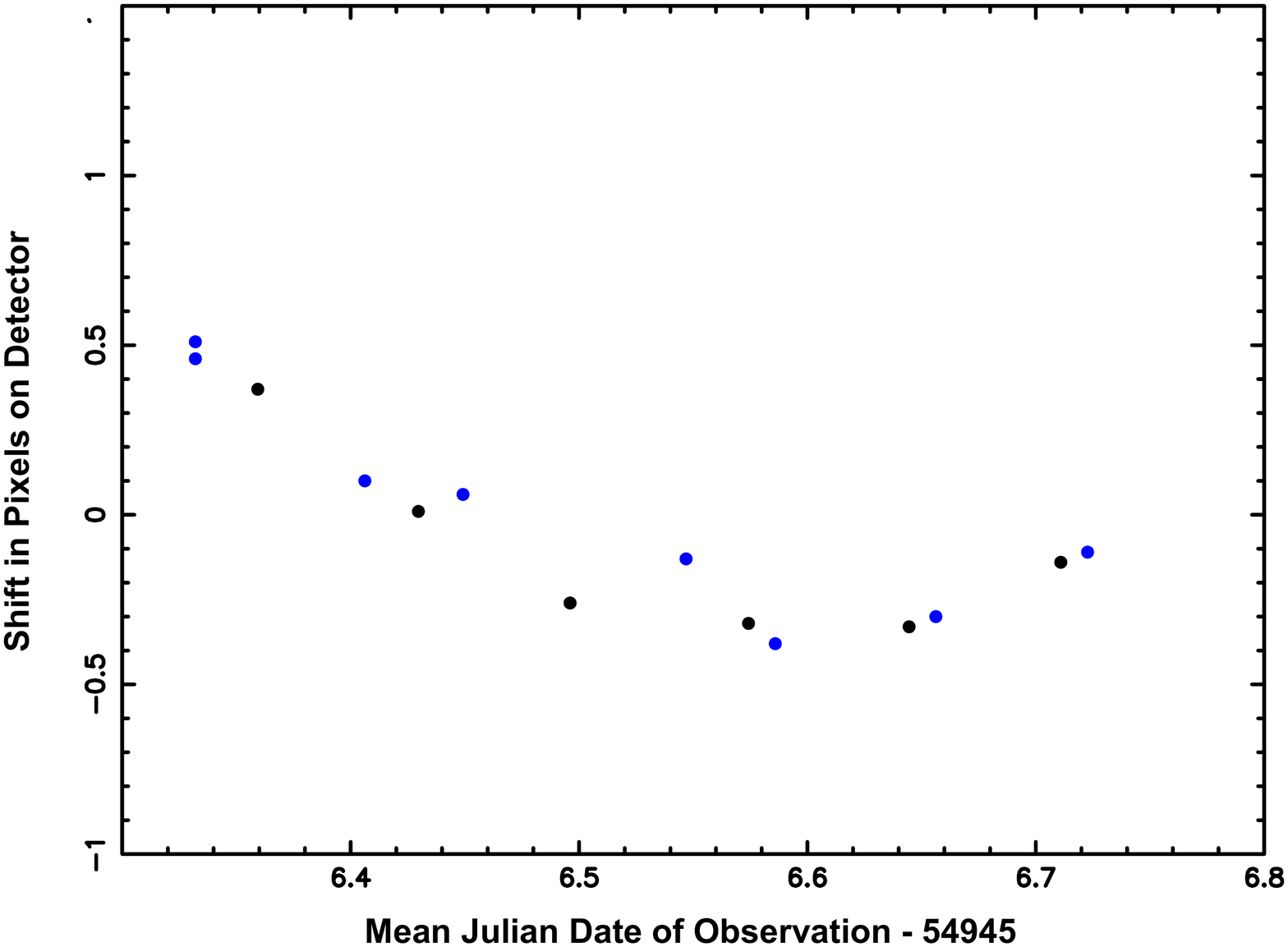}
\caption{The measured spectral drift on the detector as a function of time for an Ne-Ar arc line (blue) and a sky line (black). The drift is consistent with the initially rapid decrease in the dome temperature, which is presumably reflected in changes of the grating temperature.}\label{Drift}
\end{figure}

The measured drifts of the spectrograph are systematic with time, and correlated to some degree with the dome temperature, but not at all with the camera temperature as reported by the internal temperature sensors. This is as it should be. The measured drifts for a single night of observation are shown in figure \ref{Drift} for the red camera. The amplitude of the drift is one pixel over about four hours. This is sufficient to cause problems with sky subtraction in classical mode, but not when nod-and-shuffle modes are being employed.

\begin{figure}[t]
\centering
\includegraphics[width=0.45\textwidth]{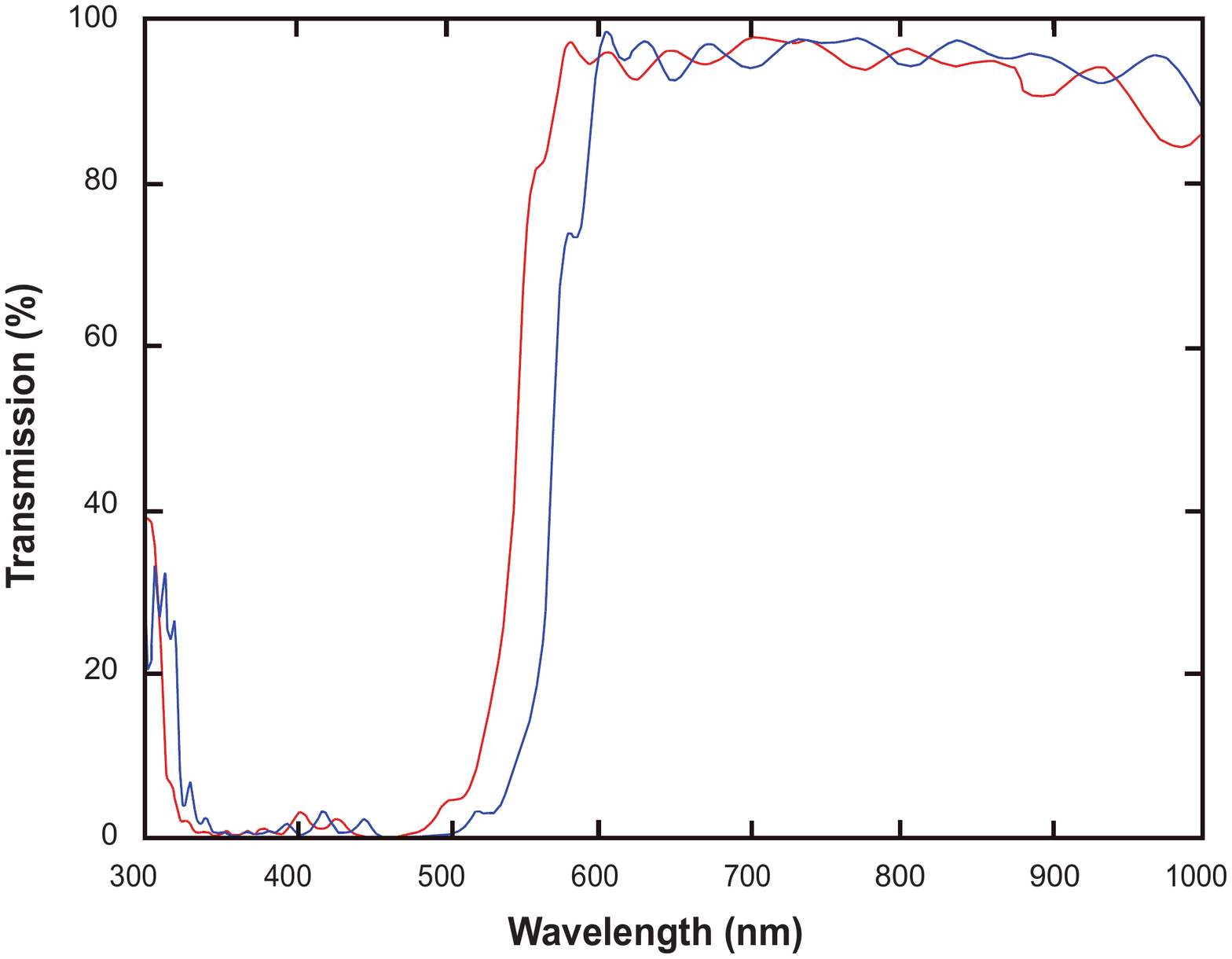}
\caption{The drift of the RT560 dichroic red transmission between the time of manufacture (red curve) and the time of deployment on the telescope (blue curve). This drift is presumably the result of annealing of the multi-layers with age.}\label{Dichroic}
\end{figure}

\subsection{Dichroic Wavelength Shift}
An unexpected effect observed during the on-telescope functional testing of WiFeS was the drift of the dichroic pass-band towards the red. This was seen in all of these Cascade Optics dichroics, and figure \ref{Dichroic} shows the effect as measured by the manufacturer on a witness sample produced along with the RT~560 dichroic. Essentially, the whole passband has moved by about 250\AA\ towards the red. This effect has ben observed in interference gratings, and is due to annealing of the multi-layers over time leading to a slight reduction in their thickness. 

Fortunately, this effect does not  seriously impact the operation of WiFeS, since the region of overlap of the B~3000 and R~3000 grating operation wide ($5300-5900$\AA) and safely encompasses the observed dichroic drift. For the other two dichroics, they are not normally used in the region of transition from blue reflection to red transmission. However, as can be seen from figure \ref{Dichroic}, there is some impact on the efficiency of science observations in the U-band.

\subsection{Detector Performance}
The readout noise of the Fairchild detectors in quad-readout mode ($\sim 5.0$ e$^{-}$) is slightly higher than was achieved in the laboratory (3.8 e$^{-}$) with single-port readout. However, the difference is not large enough to affect the science efficiency in any appreciable way. The detectors use a quad-readout through all four corner-mounted on-chip amplifiers to reduce the total read-out time with single binning to 45 sec.

The bias frames show no fixed-pattern noise, and apart from half a ``warm'' column in the red, there are no cosmetic defects in the CCDs. However, the bias shows a tendency to drift by a few electron equivalents from exposure to exposure, and it also shows some curvature at the level of about 1electron equivalent. The amount of the bias drift and structure was initially larger, but it was controlled by improving the grounding by adding a grounding strap to the telescope fork. 

To reduce the effect of bias drift in the reduction, it is recommended to take single bias frames regularly throughout the night, and to fit these by a low-order surface fit to eliminate read-out noise. No advantage was gained by taking multiple bias frames and averaging these.

Since the signal is being read out, amplified and transmitted to the data storage location by four on-chip amplifiers at the same time, the possibility arises of electrical cross-talk between the amplifiers. Since charge is bring moved upwards and downwards from the centre of the chip, and then out along the bus at the upper and lower edges of the chip, the cross talk manifests itself as false counts from a feature in the upper right quadrant of the detector showing up as a mirror image in the upper left quadrant. Similarly, features in the upper left ca transfer to the upper right, from the lower left to the lower right, and from the lower right to the lower left. This cross-talk may manifest itself both as a positive or as a negative false signal.

\begin{figure}[t]
\centering
\includegraphics[width=0.46\textwidth]{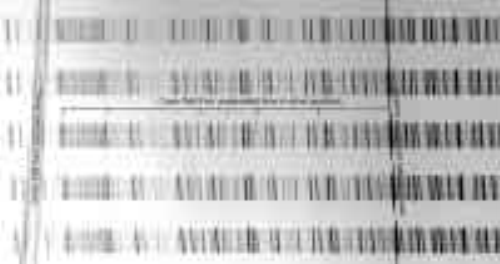}
\caption{A small portion of an arc observation (made in the coronagraphic aperture) showing both column overfill from saturation, and the effect of cross-talk between adjacent amplifiers in the read-out. Both cross-talk from unsaturated strong lines in the other quadrant, and cross-talk from very strong lines which have led to ``bleeding'' in the column direction are shown.  The contrast in this data has been turned up very high to bring out these faint features.}\label{Cross-talk}
\end{figure}

Cross-talk is particularly noticeable when the signal in one of the quadrants is saturated and has overfilled its column and has bled in the vertical direction. An example of cross-talk between quadrants is shown in figure \ref{Cross-talk}. Where the signal from the other quadrant is saturated, or is near to the saturation we detect cross-talk at about about 150 counts per pixel. However, the spatial displacement (caused by different timing of the readout cycles) of the cross-talk signal, and their mirror tilt in the spectral direction makes these cross-talk artfacts very easy to distinguish from real features of the spectrum. 

\subsection{Ghost Images}
Ghost images are potentially a problem in axially symmetric systems such as the WiFeS cameras. The ghost images that can be formed are of two types:
\begin{enumerate}
\item{Pupil image ghosts, formed in the centre of the field when reflections from lens surfaces form a roughly-focussed image of the pupil on the detector, and }
\item{Field image ghosts produced when reflections from lens surfaces form a roughly-focussed image of the field on the detector}
\end{enumerate}

The WiFeS spectrograph was designed to provide ghost image intensities which peak at less than $10^{-4}$ of the intensity in the principal spectrum. This renders any ghost image intensity comparable or less than the readout noise when the CCD is exposed to its full-well capababilty. Such faint ghost images are achieved by use of VPH gratings, very high levels of baffling against scattered light, careful optical design, and good-quality anti-reflection coatings on all air-glass surfaces. This design effort was successful, and no field or pupil ghost has been detected in the data.

However, in addition to normal ghost images, a zero-order ghost image is formed in spectrographs when VPH gratings are used in Littrow configuration. In this configuration, dispersed light passing through the camera optics can be reflected from the surface of the CCD, pass back through the optics of the camera, and then de-dispersed and sent back from the grating to form a faint white light image of the object being observed near the centre of the field. This is called a ``Littrow Ghost'', or a ``recombination ghost'' and its discovery in volume phase holographic applications has been described by \citet{burgh07}.  These authors observe this ghost in two spectrographs built by the University of Wisconsin - Madison: the Robert Stobie Spectrograph for the Southern African Large Telescope, and the Bench Spectrograph for the WIYN 3.5 m telescope. It has also been seen in relation to the AAOmega spectrograph, and discussed in an internal technical paper.

Littrow ghosts can be avoided by tilting the grating  to a non-Littrow configuration to displace the ghost image out of the detector surface. This would have required considerable change to the grating specification and a loss of efficiency, so this was not done in the WiFeS design.

The detailed theory of the Littrow ghosts has been developed by \citet{burgh07}. In summary, there are two possible explanations for the formation of Littrow ghosts:
\begin{enumerate}
\item{Light reflected from the detector is re-collimated by the camera, un-dispersed as it is transmitted back though the grating, and partially reflected by the front face of the grating substrate. For Littrow configuration, this white ghost beam approaching the grating VPH layer is parallel to the original central-colour beam leaving the grating layer. Any leakage into zero-order therefore forms a white-light ghost image of the slit (or the star being observed) across the centre of the detector.}
\item{Light reflected from the detector is re-collimated by the camera and un-dispersed as it is partially reflected off the grating. For Littrow configuration, this white ghost beam returns to the camera parallel to the original central colour beam, and so also forms a white-light ghost image of the slit or the star being observed close to the centre of the detector.}
\end{enumerate}
The latter mechanism presumably produces much the stronger ghost image since the former is attenuated by both leakage into zero-order diffraction and reflection off an AR-coated glass face.
\begin{figure}[t]
\centering
\includegraphics[width=0.3\textwidth]{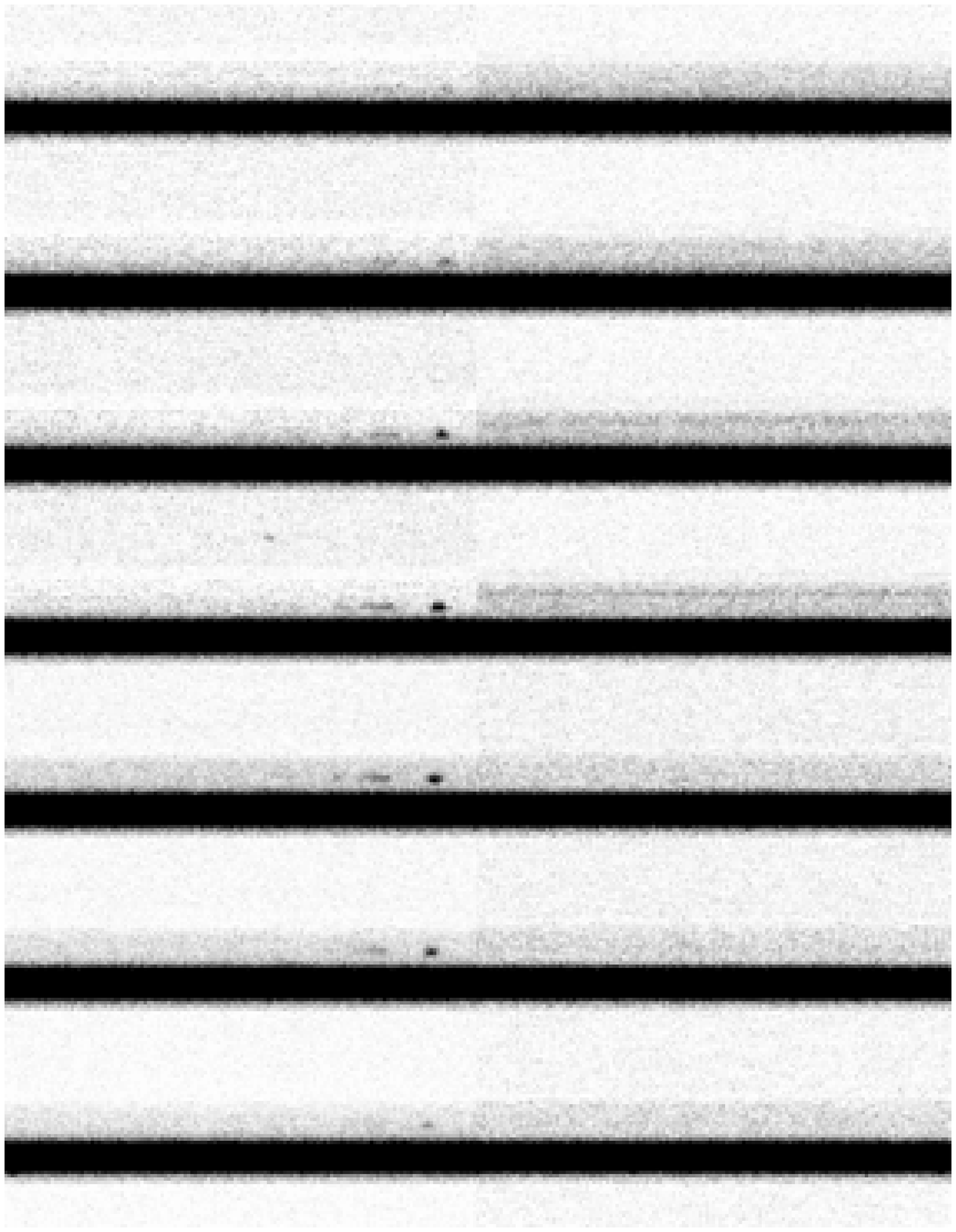}
\caption{A small portion of a standard star observation with the R7000 grating near the centre of the field. The Littrow ghost images of the star are seen in a number of slices above the stellar spectrum. The counts in the Littrow ghost show that it has an intensity $\sim 3\times 10^{-5}$ of the primary spectrum.}\label{Littrow-ghost}
\end{figure}

Figure \ref{Littrow-ghost} shows the observation of both kinds of Littrow ghosts in an observation made of a standard star under fairly poor conditions of seeing, so that the star straddles several slices of the image slicer. The ghost is displaced from the spectrum of the star because the star was displaced from the centre of the slit. Note that the primary ghost, generated by the second mechanism above, forms a white light image of the star as seen within each slice of the image. The ratio between the total number of counts in the ghost image and the total number of counts in the spectrum is measured to be $2.7 \times 10^{-5}$. This is somewhat better than the $\sim 10^{-4}$ measured by \citet{burgh07}, and it presents no important issue in the day-to-day operation of the WiFeS spectrograph.

\subsection{Throughput}
During the commissioning period, in April 2009, four photometric  nights of observation were devoted in part to the measurement of the absolute throughput of the WiFeS instrument. This was done by measuring the number of counts per \AA\ produced by the standard LTT~4364 and HD128279 observed in the classical observational mode with all the standard sets of gratings and dichroics. The exposure times were 600 sec. 

The small-scale (high frequency) pixel-to-pixel sensitivity variations mostly produced by fringing in the detector in the red, or by variations in the CCD doping in the blue, were removed by use of flat field observations made with the internal quartz iodine lamps. The large-scale variation across the detector was approximated by a Chebyshev Polynomial surface fit, and the flat field was divided into this surface fit so as to produce a map of the high frequency sensitivity fluctuations. This function has a mean of unity, so on average does not affect the observed number of counts from the standard star. In addition to this correction, the telluric absorption features in the red were removed using observations of a B-dwarf type star telluric standard.

\begin{figure*}[t]
\centering
\includegraphics[width=0.8\textwidth]{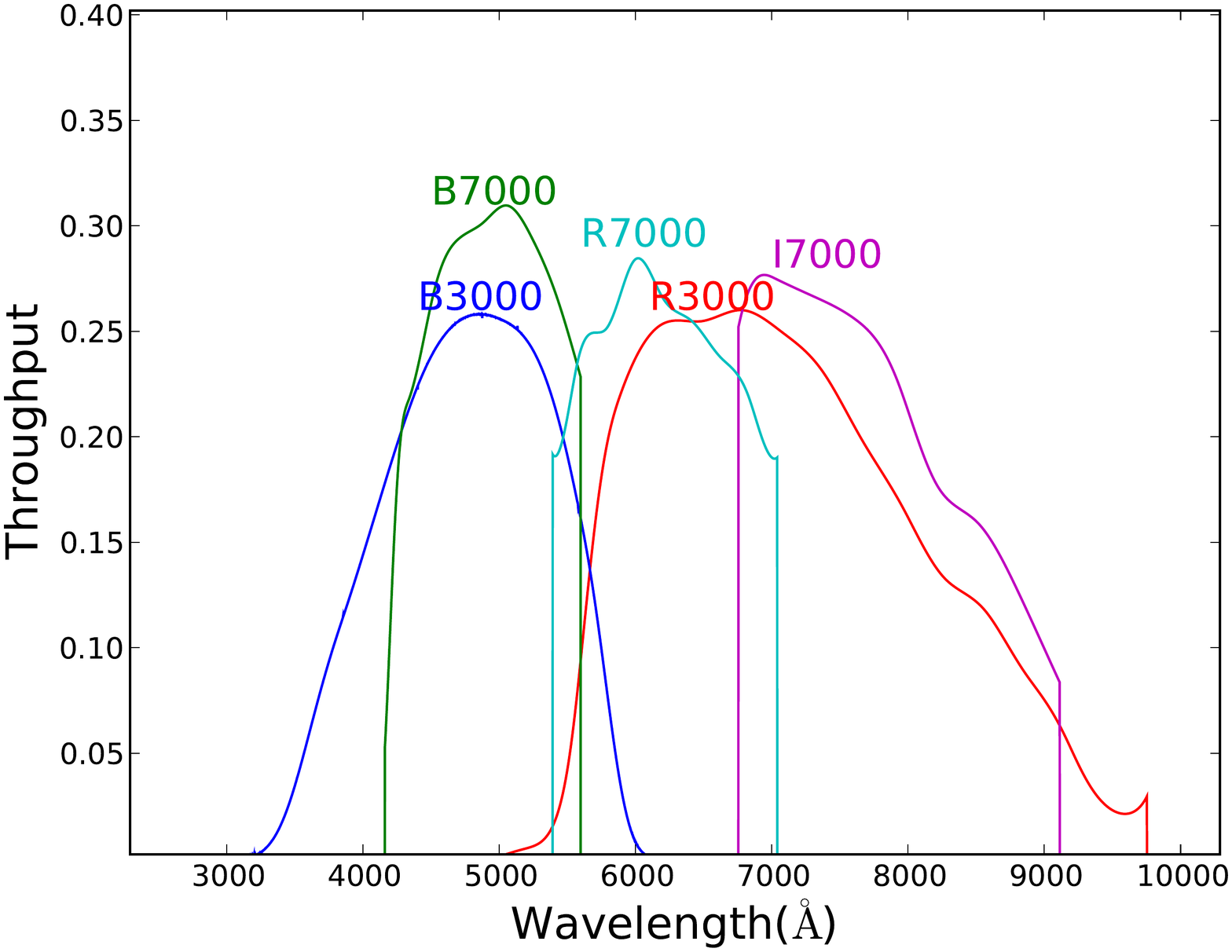}
\caption{Measured throughput Curves of the WiFeS Grating and Dichroic sets. The apparent increase at the long-wavelength limit of the R3000 grating is probably an artefact of the reduction procedure. These curves include both the transmission of the atmosphere and of the telescope.}\label{throughput}
\end{figure*}

It was not possible to determine the throughput for the U~7000 \& RT~480 grating/dichroic combination, as the flat fields did not contain enough counts to allow the reduction program to trace and extract he spectra, and furthermore, the wavelength calibration could not be reliably established. The throughput so determined for the other grating and dichroic combinations is shown in figure \ref{throughput}. This throughput is end-to-end and so includes the transmission of the telescope and of the atmosphere as well as the quantum efficiency of the detectors and the transmission of the spectrograph. 

Figure \ref{throughput} should be compared with Figure 12 of \citet{dop07}. It will be noted that, except for the B~3000 grating, the on-telescope sensitivity is in general only 80\% of what had been computed from a component by component analysis, and only about 70\% of what had been expected in the case of the R~3000 grating. There was an issue with the re-aluminisation of the primary mirror (possibly during cleaning) which led to the reflective film becoming crazed in a few regions. However, this could account for $\sim 5$\% at most. The low efficiency of the R~3000 grating and the relatively high efficiency of the B~7000 grating suggests that the main issue lies with the gratings themselves, either due to errors in the laboratory measurements of their throughput. Indeed, the throughput estimates obtained at the time of the Critical Design Review (CDR) are in much better agreement with the measured performance figures. The peak transmission and wavelength of peak transmissions estimated at the time of the CDR are here compared with the measured throughputs and peak wavelengths at peak (given in parentheses): B~3000, 32\%, 4200\AA\ (26\%, 4800\AA); B~7000, 34\%, 4900\AA\ (32\%, 4900\AA); R~7000, 31\%, 6200\AA\ (28\%, 6000\AA); R~3000, 30\%, 7100\AA\ (26\%, 6900\AA); and I~7000, 32\%, 8000\AA\ (28\%, 7000\AA). Since the transmissions of the VPH  gratings at the time of the CDR were estimated from the manufacturer's figures, we may conclude that the laboratory measurements of the VPH gratings which were made subsequently were in error, presumably because of fundamental limitations in the absolute calibration, and possibly also in inadequate order-sorting.

With its throughput as measured, WiFeS is an extraordinarily efficient spectrograph offering greater than 20\% throughput between 4000 and 8000\AA. Allowing for the telescope, atmosphere and detector, the throughput of the spectrograph by itself peaks at over 45\%. We are not aware of an integral field spectrograph anywhere else in the world which matches the combination of throughput, wavelength and field coverage offered by WiFeS. Indeed, the reported throughput of SPIRAL, an integral field unit placed on the AAOmega spectrograph of the Anglo-Australian Telescope is only 12\% at 7000\AA\ (\emph{vs.} 42\% for WiFeS) \citep{green09}. This instrument is the only other integral field unit available on an Australian telescope. Compared with the Double Beam Spectrograph (DBS) \citep{Rodgers88} which it replaces, WiFeS is capable of accumulating data on extended objects at a rate of $\sim 200$ times faster, and on stellar object at a rate of  $\sim 6$ times faster.

\section{Sensitivity}
The sensitivity of the instrument to either point or extended sources can be readily computed from the throughput, the night sky brightness and the read-out noise and dark counts of the CCD. The read out noise and dark counts of the CCD are listed above in Table \ref{Table2}. The night sky brightness as a function of wavelength is taken from \citet{benn07}. 

Based upon these figures, an web-based exposure time calculator for general observers has been placed on the WiFeS web page: \textsf {http://www.mso.anu.edu.au/ \newline observing/ssowiki/index.php/WiFeS\_Main\_Page}.

Assuming a seeing of 1.0 arc sec, the computed limiting magnitudes for stars (to deliver a signal to noise ratio of 3 per resolution element in an exposure time of 1 hour split into three separate exposures in order to remove cosmic rays are given in Table \ref{Table3}.

\begin{table*}[t]\label{Table3}
\centering
\caption{Limiting Magnitudes for WiFeS (to reach S/N=3 per resolution element in a 3$\times$1200s exposure).}
\begin{tabular}{@{}lcccccc@{}}
\tableline \tableline 
& & {\bf Blue} & & & {\bf Red} & \\
& \multicolumn{2}{c}{$R = 7000$} & {$R = 3000$} & \multicolumn{2}{c}{$R = 7000$} & {$R = 3000$} \\
\tableline
Grating  & U7000 & B7000 & B3000 & R7000 & I7000 & R3000  \\
Dichroic & RT480 & RT615 & RT560 & RT480 & RT615 & RT560  \\
$\lambda_{eff}$ (\AA) & 3850 & 4900 & 4680 & 6200 & 8000 & 7420  \\
Mag. & 20.00 & 21.55 & 22.30 & 21.40 & 20.90 & 22.05  \\
\tableline

\end{tabular}
\end{table*}

\begin{figure*}[t]
\centering
\includegraphics[width=0.8\textwidth]{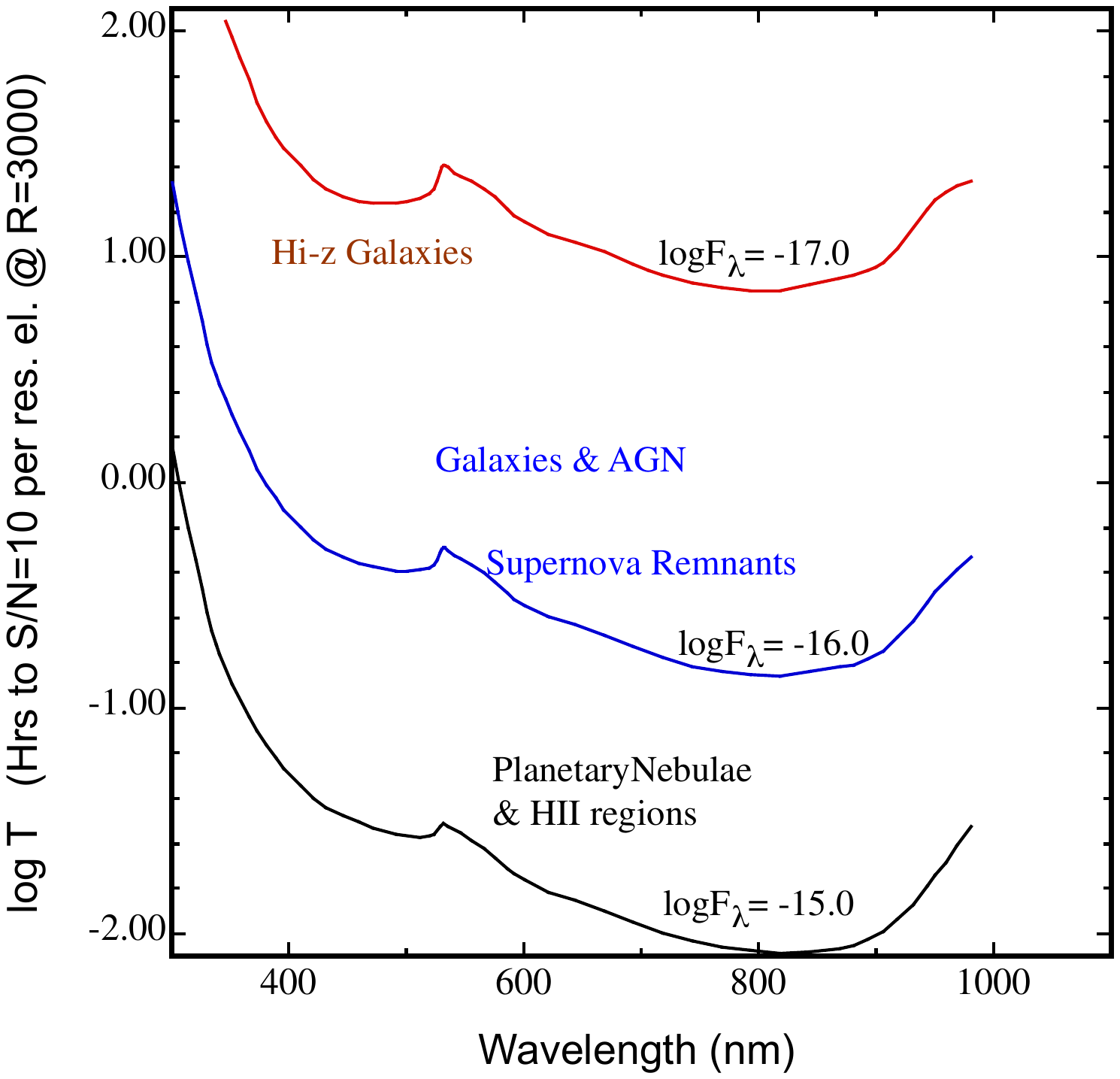}
\caption{The exposure time in hours needed for WiFeS to reach a signal to noise of 10 per resolution element at a resolution of $R=3000$ for three values of the surface brightness (ergs cm$^{-2}$ s$^{-1}$arcsec$^{-2}$\AA$^{-1}$). The approximate regimes characteristic of different classes of extended astronomical targets are indicated on the figure.}\label{sensitivity}
\end{figure*}

For extended objects such as galaxies or emission line objects, it is more convenient to express the sensitivity in term of the surface flux, $F_{\lambda}$, measured in ergs cm$^{-2}$ s$^{-1}$arcsec$^{-2}$\AA$^{-1}$. The exposure time required to reach a signal to noise ratio of 3 in each resolution element at a resolution of 3000 is given in Figure \ref{sensitivity} for three different surface fluxes.

\section{Conclusions}
The functional testing of the WiFeS spectrograph has revealed that it achieves all the performance requirements defined by its science mission. Its optical performance surpasses the requirements, except in respect of the throughput, which is slightly below expectations. Only minor cosmetic issues have been encountered. These relate to Littrow ghosts and cross-talk between the on-chip amplifiers in quad-readout mode.

All modes of operation have been implemented through the Telescope and Remote Observing software, and the remote observing capability itself has been tested and demonstrated. Finally, the data reduction pipeline has been implemented and proven fit for purpose, although it continues to be enhanced in the light of operational experience.

The WiFeS instrument has already revived the demand for ANU 2.3m telescope time, and the over-subscription rate now approaches three. It has become the main research tool for a number of student theses -  a key objective of the Australian Department of Science and Education (DEST) Systemic Infrastructure Initiative grant which made its construction possible. With the advent of WiFeS, the 2.3m telescope has once more been transformed into a fully-competitive major research facility.

\begin{acknowledgments}
The authors of this paper wish to profoundly thank all members of the engineering team, current and past,  who have made possible the WiFeS instrument. They also acknowledge the support of the Australian Department of Science and Education (DEST) Systemic Infrastructure Initiative grant which provided the major funding for this project. 

They also acknowledge the receipt of an Australian Research Council (ARC) Large Equipment Infrastructure Fund (LIEF) grant LE0775546, which allowed the construction of the blue camera and associated detector module. Dopita acknowledges the continued support of the Australian Research Council (ARC) through Discovery  projects DP0984657 and DP0664434.  Bussarello,  Merluzzi \& Dopita acknowledge travel support under the European Union FP7 Marie Curie Actions International Research Staff Exchange Scheme (IRSES).
\end{acknowledgments}


\end{document}